# Seeding the $m = 0$ instability in dense plasma focus Z-pinches with a hollow anode

J. X. Liu,[1,2] J. Sears,[1] M. McMahon,[1] K. Tummel,[1] C. Cooper,[1] D. Higginson,[1] B. Shaw,[1] A. Povilus,[1] A. Link,[1] and A. Schmidt[1]

[1]Lawrence Livermore National Lab, Livermore, CA 94550, USA

[2]Department of Physics, UC Berkeley, Berkeley, CA 94720, USA

The dense plasma focus (DPF) is a classic Z-pinch plasma device that has been studied for decades as a radiation source. The formation of the $m = 0$ plasma instability during the compression phase is linked to the generation of high-energy charged particle beams, which, when operated in deuterium, lead to beam-target fusion reactions and the generation of neutron yield. In this paper, we present a technique of seeding the $m = 0$ instability by employing a hollow in the anode. As the plasma sheath moves along the anode's hollow structure, a low density perturbation is formed and this creates a non-uniform plasma column which is highly unstable. Dynamics of the low density perturbation and preferential seeding of the $m = 0$ instability were studied in detail with fully kinetic plasma simulations performed in the Large Scale Plasma particle-in-cell code as well as with a simple snowplow model. The simulations showed that by employing an anode geometry with appropriate inner hollow radius, the neutron yield of the DPF is significantly improved and low-yield shots are eliminated.

## 1. INTRODUCTION

The dense plasma focus (DPF) is a classic plasma device consisting of two coaxial, cylindrical electrodes separated at one end by an insulator and filled with a low pressure gas [1]. The inner electrode is the anode and the outer electrode is the cathode. The dynamics of the discharge can be divided into four main phases, illustrated in Fig. 1. First is the breakdown phase, in which a high voltage pulse applied between the electrodes ionizes the gas above the insulator to form a plasma sheath. Subsequently, the $\mathbf{J} \times \mathbf{B}$ force lifts the sheath off the insulator and drives it down the axis of the DPF in the axial run-down phase. In this phase, the neutral background gas is ionized and swept up by the magnetic field, increasing the sheath's mass and density. This process also yields a kinetic ram pressure that is dependent on the sheath velocity and counteracts the $\mathbf{J} \times \mathbf{B}$ force [2]. Once the sheath has reached the axial end of the anode, the radial implosion phase begins. In this phase, the axially travelling portion of the sheath continues on its trajectory, but a new region of plasma sheath forms – still connected to the axial sheath – and implodes in the radial direction, similarly driven by the magnetic field pressure and counteracted by the ram pressure. This radially travelling section of the sheath also ionizes and accrues mass from the background gas as it travels, but it begins this process anew: the mass swept up in the axial portion of the sheath does not transfer to the radial segment. Last is the pinch phase, which occurs when the plasma sheath collides on-axis to form a hot and dense column. During this phase, various instabilities develop and break apart the plasma [3].

One instability in particular, the $m = 0$ "sausage" mode, is regarded as an important factor in the formation of high energy particle beams and, when operated with deuterium or deuterium-tritium gas, the emission of fusion neutrons from the dense plasma focus. The $m = 0$ instability necks and subsequently severs the plasma column, generating intense electric fields in the cavity between the two separated portions of the plasma [4]. This axial electric field accelerates ions into the high density plasma and background gas, yielding beam-target fusion neutrons [5]. In low-energy dense plasma foci machines, a significant portion of the neutron yield is expected to constitute of beam-target fusion due to the divergence of the measured yields from those predicted by $\propto I_{peak}^4$ thermonuclear scaling models [6, 7, 8].

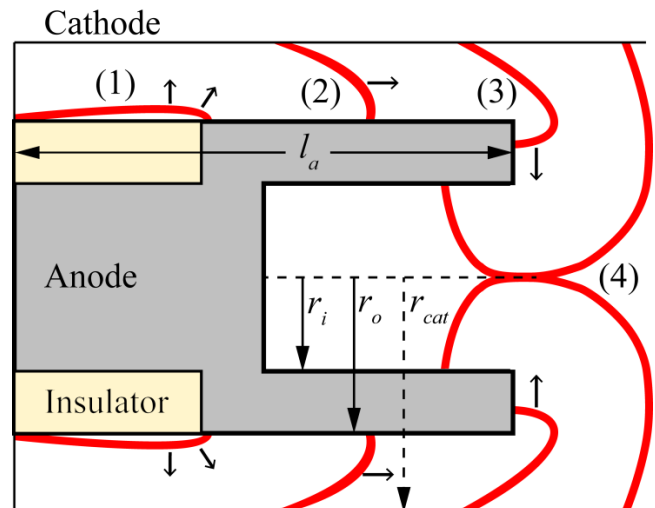

FIG. 1: Diagram of the DPF geometry and of the plasma sheath, in red, as it propagates at various times throughout the discharge. The numbers refer to: (1) breakdown along the insulator, (2) axial run-down phase, (3) radial implosion phase, and (4) pinch phase. The black arrows indicate the direction of sheath propagation.

## 2. LOW DENSITY SHEATH FORMATION

When the plasma sheath has reached the end of the anode in the axial phase, as illustrated in Fig. 2 (a), the plasma has a large axial component of momentum and essentially no radial component. Therefore, upon reaching the edge of the anode, the bulk of the plasma, represented by the red section in Fig. 2 (b), continues on its axial trajectory. To maintain a conductive path to the anode, current must now flow through background density gas, forming the radially imploding section of the sheath illustrated by the green region. Initially, this radial section of the sheath has a significantly lower density than the rest of the plasma, although this lasts only for several tens of nanoseconds. As this section propagates radially, it ionizes and sweeps up the background gas, rising to a density comparable to the remainder of the sheath, as shown in Fig. 2 (c).

In a typical DPF with a completely solid anode, the radial segment of the sheath continually grows denser until it collides on-axis in a hot, dense column. However, if a hollow of radius $r_i$ is introduced to the anode, it is possible to create yet another low density perturbation to the plasma during the implosion phase. When the sheath reaches $r_i$, the original, high density (HD) branch of the radial segment continues on its radial trajectory, as illustrated in Fig. 2 (d) in red. Just as in the case of reaching the axial end of the anode, a discharge path is created through the background gas, and a low density (LD) branch is formed within the anode hollow, indicated by the green section. The newly formed LD branch is lower in density than the HD branch, yet it carries the same discharge current and experiences the same magnetic field. Therefore the LD branch experiences a greater radial acceleration and its trajectory diverges from that of the HD branch. This difference can be exploited by choosing an optimal hollow radius to create an axially non-uniform plasma column that preferentially seeds the $m = 0$ instability. These dynamics and their implications on the beam production and neutron yield of the DPF were studied in detail with numerical simulations in the particle-in-cell code Large Scale Plasma (LSP), as well as with a simple snowplow model.

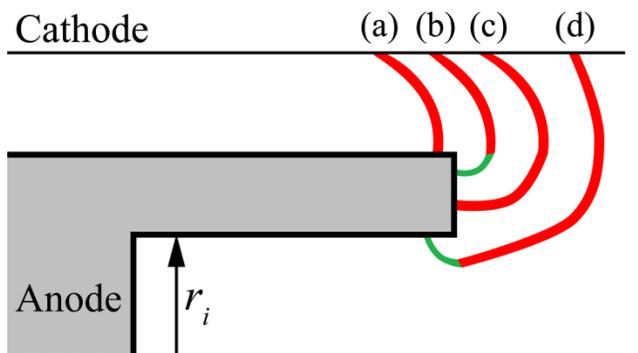

FIG. 2: Diagram of the low density sheath formation at the axial end of the anode and at the anode's inner hollow. The red regions of the sheath correspond to locations of high density, while the green regions correspond to the newly formed locations of low density. As the sheath propagates, the low density regions sweep up mass and become denser. Only the upper half of the cross sectional image is shown.

## 3. DENSE PLASMA FOCUS GEOMETRY

The DPF geometry used in this study was adapted from the Nanofocus device, developed by M. Milanese, et al. [6]. The parameters and dimensions used in this study are similar to those of the Nanofocus. The presently studied device has a charging voltage of 16 kV, total inductance of 58.7 nH, and stored bank energy of 140 J. The peak current is 62 kA. The anode's outer radius is 0.75 cm and its length is 2.4 cm; the cathode's outer radius is 2.15 cm. The device operates with a background pressure of 200 Pa of molecular deuterium.

This particular geometry was chosen because experimental studies by Milanese demonstrated neutron yield scaling which was well above the conventional $\propto I_{peak}^4$ thermonuclear scaling law, suggesting a significant beam-target fusion yield, which the low density sheath formation attempts to enhance by seeding the $m = 0$ instability.

This geometry is also technologically interesting because its low stored energy and small dimensions (fitting within a cube 40 cm on a side) make this device very portable. This mobility is essential to the potential deployment of such a device as a portable neutron source for various applications such as oil well logging [9] or special nuclear material detection [10].

## 4. NUMERICAL SIMULATION IN LSP

The dense plasma focus is numerically simulated with the particle-in-cell code LSP [11]. The simulation geometry is two-dimensional in cylindrical coordinates ($r$, $z$), as the simulation is symmetric about the $\theta$ direction. A grid with 220 cells in the $r$ direction and 300 cells in the $z$ direction covers a radial extent of 1.075 cm and an axial extent of 6 cm. LSP employs an implicit algorithm to simultaneously push particles and calculate electromagnetic fields on a grid. This allows for long time steps to be taken and under-resolution of $\lambda_{Debye}$ [12]. The algorithm operates in the regime where $\omega_c \Delta t < 0.3$ is fulfilled, where $\omega_c$ is the electron cyclotron frequency and $\Delta t$ is the simulation time step. The discharge is simulated in two phases: an MHD fluid phase followed by a fully kinetic phase. The simulation time step begins at $1.0 \times 10^{-2}$ ns during the MHD fluid phase and is eventually ramped down to $2.0 \times 10^{-4}$ ns during the kinetic phase. The plasma is initiated as a MHD fluid to reduce computational time. When the plasma has reached sufficient proximity to the axis, then the code is switched into a fully kinetic simulation to capture the relevant physics during the pinch, including beam formation [12]. The background gas is initialized as fully ionized plasma at room temperature with a 0.5 mm thick, 1 eV sheath initialized over the insulator to emulate the plasma after breakdown.

## 5. SEEDING THE $m = 0$ INSTABILITY

At the beginning of the simulation, the capacitor bank is switched onto the electrodes and current flows through the high temperature sheath. The $\mathbf{J} \times \mathbf{B}$ force lifts the plasma up off the insulator and accelerates it down the axis of the DPF, reaching velocities up to $6.6 \times 10^4$ m/s. Fig. 3 shows snapshots of the sheath position at various times throughout the axial run down phase.

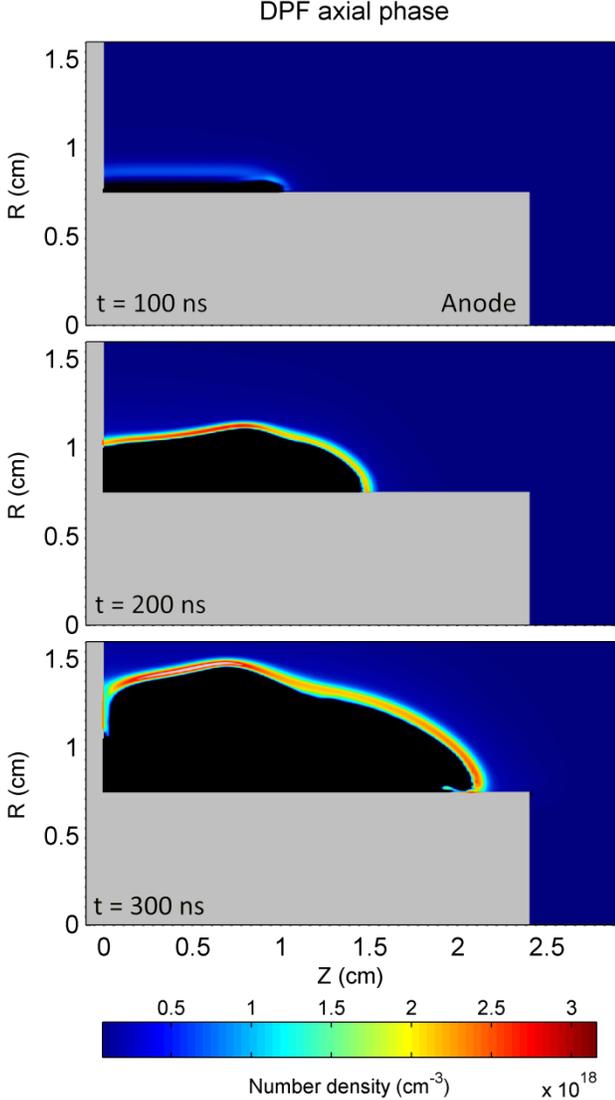

FIG. 3: Snapshots of the ion density at various times during the axial run-down phase. To preserve the dynamic range of the color scale, densities below $1 \times 10^{16}$ cm$^{-3}$ are represented by black.

When the plasma sheath reaches the axial end of the anode, the mass swept up during the axial phase continues on its original trajectory due to its acquired axial momentum. At this time, the radial section of the sheath forms and begins propagating towards the axis. The trajectory of this radial segment is shown in the blue curve of Fig. 4. Initially, this segment has a low density, comparable to the background density, although this does not last. As the sheath propagates, it sweeps up background gas, and the temporal evolution of the sheath's density is shown in the green curve of Fig. 4.

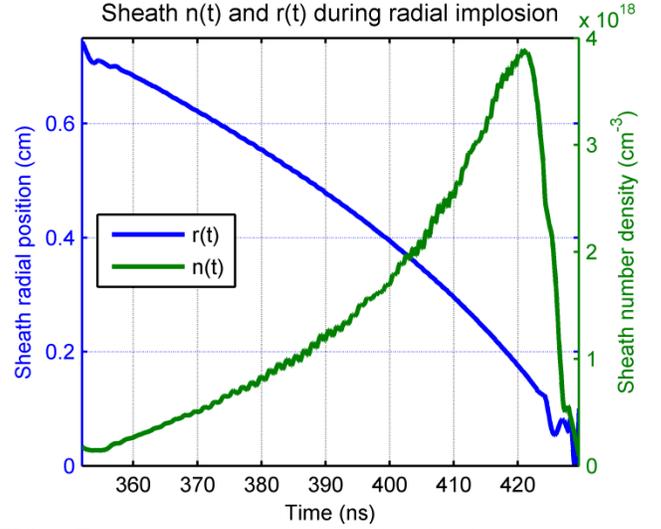

FIG. 4: The sheath's radial position and number density versus time, during the radial implosion. The periodic oscillations seen in the plots of $r(t)$ and $n(t)$ are caused by the discrete mesh size of the simulation.

To study the low density sheath formation and its application to preferentially seed the $m = 0$ instability, anode inner radii of $r_i = 0.1875$ cm, $0.3750$ cm, $0.5625$ cm, and $0.7200$ cm were simulated, all with the same outer radius, $r_o = 0.7500$ cm, and with all other aspects of the simulations kept the same. Snapshots of the ion density at various times throughout the radial implosion are shown in Fig. 5.

Fig. 5 (a) shows the radial implosion with the anode without a hollow. All throughout the implosion, the plasma sheath assumes a relatively smooth density profile, with little variation along the axial direction. After reaching the axis in Fig. 5 (a, 4), instabilities eventually develop and evacuate the cavity seen in Fig. 5 (a, 5). The cavity reaches a width of 0.4 mm and a density of $3.0 \times 10^{17}$ cm$^{-3}$. The hollow is introduced in Fig. 5 (b) at $r_i = 0.1875$ cm. In this simulation, the low density region reaches the axis several nanoseconds in advance of the bulk of the sheath, creating an axially non-uniform column. In this column, an $m = 0$ instability is seen to develop in the region where the low density sheath first reached the axis, severing the column into two parts. A cavity as wide as 1.2 mm forms and it reaches densities as low as $3.3 \times 10^{16}$ cm$^{-3}$. The anode with inner hollow $r_i = 0.3750$ cm performs similarly, generating a cavity of width 0.8 mm and density $3.6 \times 10^{16}$ cm$^{-3}$. The last two anodes, with hollow radii 0.5625 cm and 0.7200 cm, perform worse. Although the low density sheath does form in both cases, it rises to a density comparable to the high density branch. During the onset of instabilities, as seen in Fig. 5 (d, 4) and Fig. 5 (e, 5), the cavity width reaches 0.4 mm and 0.2 mm, respectively, and the density reaches $1.5 \times 10^{17}$ cm$^{-3}$ and $1.3 \times 10^{17}$ cm$^{-3}$, respectively.

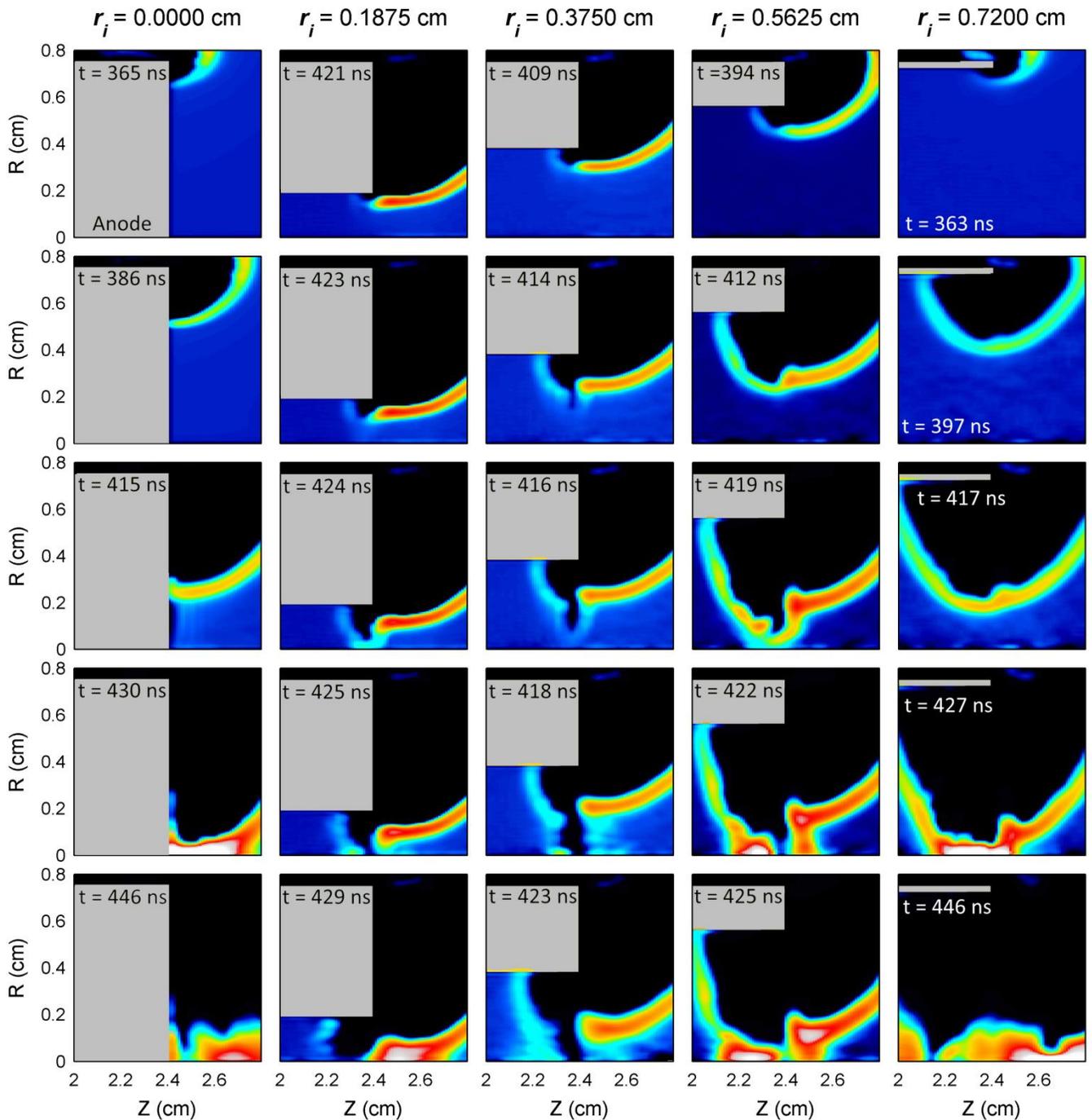

FIG. 5: Snapshots of the simulated ion densities at various times during the radial implosion phase for the various inner radii. From left to right, (a) through (e) refers to the column while from top to bottom, (1) through (5) refers to the row. The simulation time is shown for each frame.

## 6. ENHANCEMENT OF NEUTRON YIELD

In this device, the $m = 0$ instability is regarded as an important factor in the generation of the electric fields that accelerate deuterons to high energies [4, 13, 6]. Therefore, it is expected that promoting the formation of the instability is beneficial to the overall neutron yield of this DPF.

The use of a hollow in the anode creates a perturbation in the radially imploding plasma annulus. The low density portion of the annulus accelerates towards the axis and collides on-axis before the high density region of the sheath has arrived. This creates an azimuthally symmetric but axially non-uniform plasma column – ideal for the sausage instability to develop and generate strong axial electric fields. However, this effect only occurs effectively for a particular range of inner radii. If the hollow radius is too large, then the LD branch comes into existence relatively early and it has sufficient time to build up to a density comparable to the rest of the sheath. In such a case, the result is yet again a relatively uniform z-pinch column that does not preferentially seed an instability.

The various inner radii were simulated in seven identical runs, and their neutron yields are shown in Fig. 6. The average yield is significantly greater for the smallest inner anode radius, with over 4x improvement seen between $r_i = 0.1875$ cm and $0.7200$ cm. Additionally, for the larger hollow radii, $0.5625$ cm and $0.7200$ cm, there are many shots which yield below $1 \times 10^6$ neutrons. These are interpreted as shots in which no substantial instability developed and no intense electric fields were generated. The smaller inner radii, $0.1875$ cm and $0.3750$ cm, seem to have eliminated these "dropout" shots, and this is vital for applications that require a reliable yield.

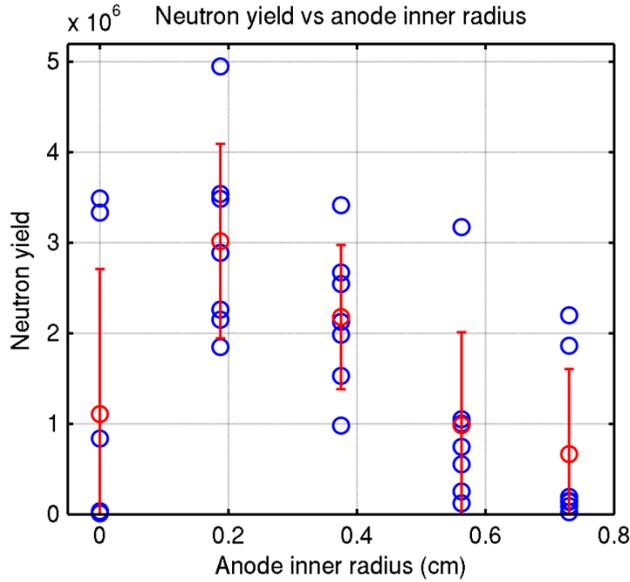

FIG. 6: The neutron yields of the various simulated anode geometries. The inner radii include $r_i = 0.0000$ cm (no hollow), $0.1875$ cm, $0.3750$ cm, $0.5625$ cm, and $0.7200$ cm. Seven identical runs were simulated at each $r_i$, and their yields are shown as blue circles. The red circles indicate the average yield, and the red bars indicate the standard deviation.

Since the neutron yield of this device is expected to primarily stem from beam-target fusion reactions, the different yields are explained by the difference in the generated ion beams. The ability of a given beam to produce neutrons is characterized by its energy distribution function, $f(E)$, weighted by the deuterium-deuterium fusion cross section, $\sigma(E)$.

Fig. 7 shows this weighted distribution during the time of peak neutron production, averaged over all simulations at each inner radius. There is greater noise in the distributions at higher energies because there are fewer simulation particles in that range.

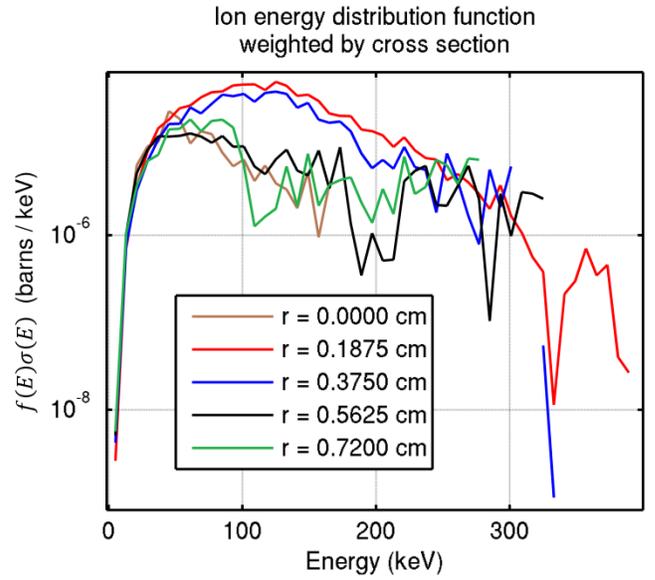

FIG. 7: $f(E)$ weighted by $\sigma(E)$ during peak neutron production, plotted versus particle energy, and averaged over all simulations at a particular radius.

Using this ion energy distribution function, we can calculate the beam's reaction rate parameter, $<\sigma v>$.

$$<\sigma v> = \int_0^\infty \sqrt{\frac{2E}{m}}\, \sigma(E)\, f(E)\, dE$$

The comparison of the reaction rate parameter between different anode inner radii is shown in Fig. 8. The greater this reaction rate is, the greater the neutron yield is expected to be for a given beam and target density. Therefore, it is reasonable to see that the average beam reaction rate is roughly ordered the same as the neutron yields of Fig. 6. By effectively seeding the $m = 0$ instability with smaller anode inner radii, a more reactive ion beam can be formed, generating greater yields.

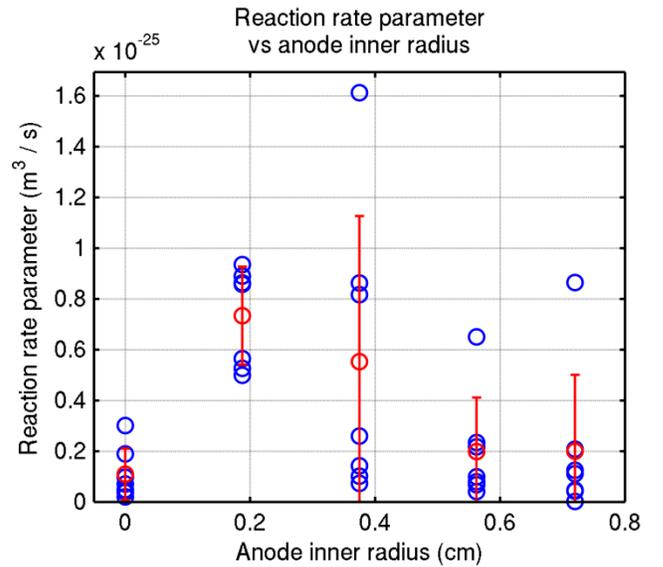

FIG. 8: Reaction rate parameter versus inner radius. Red circles indicate the average value, and red bars indicate standard deviation.

## 7. SNOWPLOW MODEL

The snowplow model [14] is applied to this DPF geometry to study the dynamics of the radial implosion phase in a simple manner. The model considers two annular slugs imploding towards the axis, illustrated in Fig. 9. The slugs are driven by the magnetic field pressure and counteracted by the dynamic gas ram pressure [2]. One slug is the high density (HD) branch of the sheath, the other is the low density (LD) branch, and the slugs are independent of each other. The equation of motion for either slug is:

$$\frac{\partial v_s}{\partial t} \rho_{slug} \Delta + v_s \frac{\partial \rho_{slug}}{\partial t} \Delta = \frac{B_\theta^2}{2\mu_0} - \rho_0 v_s^2$$

Where $\Delta$ is the radial thickness of the slug, $v_s$ is the slug velocity, $\rho_{slug}$ is the slug's mass density, $B_\theta$ is the azimuthal magnetic field at the slug's radial position calculated with Ampere's Law, $\mu_0$ is the vacuum permeability, and $\rho_0$ is the background mass density. A numeric approach is taken to solve the differential equation:

$$\frac{\partial v_{s,i}}{\partial t} = \frac{\left(\frac{B_{\theta,i}^2}{2\mu_0} - \rho_0 v_{s,i}^2 - v_{s,i} \frac{\partial \rho_{slug,i}}{\partial t} \Delta\right)}{\rho_{slug,i} \Delta}$$

$$v_{s,i+1} = v_{s,i} + \frac{\partial v_{s,i}}{\partial t} \cdot dt$$

$$r_{s,i+1} = r_{s,i} + v_{s,i} \cdot dt$$

Where the subscript $i$ refers to the present index and $i + 1$ refers to the index of the following time step. In this calculation, the initial position of the HD slug is $r_{HD} = r_o = 0.7500$ cm. The same anode inner radii are simulated, and the initial position of the LD slug is at these different inner radii: $r_{LD} = r_i = 0.1875$ cm, 0.3750 cm, 0.5625 cm, and 0.7200 cm. The LD slug only comes into existence when the HD slug has reached the position of the inner hollow: $r_{HD} = r_i$. The HD and LD slugs have constant thicknesses: $\Delta_{HD} = 0.03$ cm and $\Delta_{LD} = 0.02$ cm, and both slugs have axial length: $l = 0.10$ cm. These values were informed by simulation results. The mass sweeping factor, which determines the fractional amount of background gas that is swept into the sheath, is 0.7 for the LD branch and 0.9 for the HD branch. The background gas density is $1.0 \times 10^{17}$ cm$^{-3}$ and the deuteron mass is $3.34 \times 10^{-27}$ kg. The discharge current is approximated by a constant: $I = 62$ kA, the peak current in the LSP simulations. The time steps taken are: $dt = 1 \times 10^{-11}$ s.

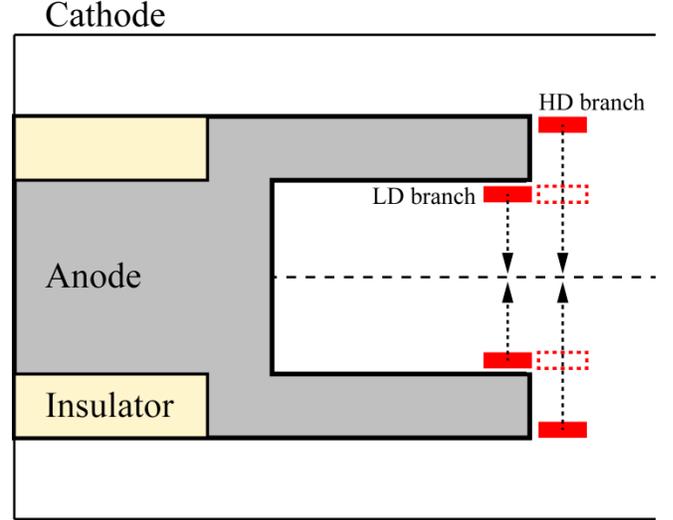

FIG. 9: Diagram of the low density (LD) and high density (HD) annular slugs in the snowplow model. The LD branch is only initiated once the HD branch reaches the "cliff" in the anode, at $r_i$.

The calculated radial trajectories of the high and low density branches are shown in the first row of Fig. 10. The time $t = 0$ ns corresponds to the instant when the HD slug passes $r_i$ and the LD branch is created. The simulated trajectories obtained from LSP are shown in the second row of Fig. 10. In LSP, the radial position of the plasma sheath is taken to be the position of peak density for a given axial position. For the LD branch, the axial position is 0.05 cm within the anode, and for the HD branch the position is the same distance outside the anode.

As soon as the LD branch comes into existence, it quickly accelerates and its trajectory diverges from that of the HD branch, although this effect is only prominent for the smaller hollow radii. For larger inner radii, the HD branch has not acquired significant mass before the LD branch is initiated, so there is less of a difference in the trajectories of the two slugs. The first row of Fig. 11 shows the snowplow model's calculated densities of the LD and HD branches during their implosion, and the second row of Fig. 11 shows the densities of the LD and HD branches simulated in LSP. Once again, $t = 0$ ns corresponds to when the HD slug passes $r_i$.

The trajectories of the slugs predicted by the snowplow model match well with those of the LSP model, differing in their times to reach the axis only by several ns or less. Additionally, the temporal evolution of the sheath densities is well-recreated by the snowplow model, suggesting that the snowplow model can be used as a simple tool to estimate the radial trajectories and densities of the LD and HD branches. This demonstrates that the mechanism responsible for the improvement in neutron yield is captured by a 1D MHD model, offering us a computationally inexpensive tool for determining an optimal anode inner radius.

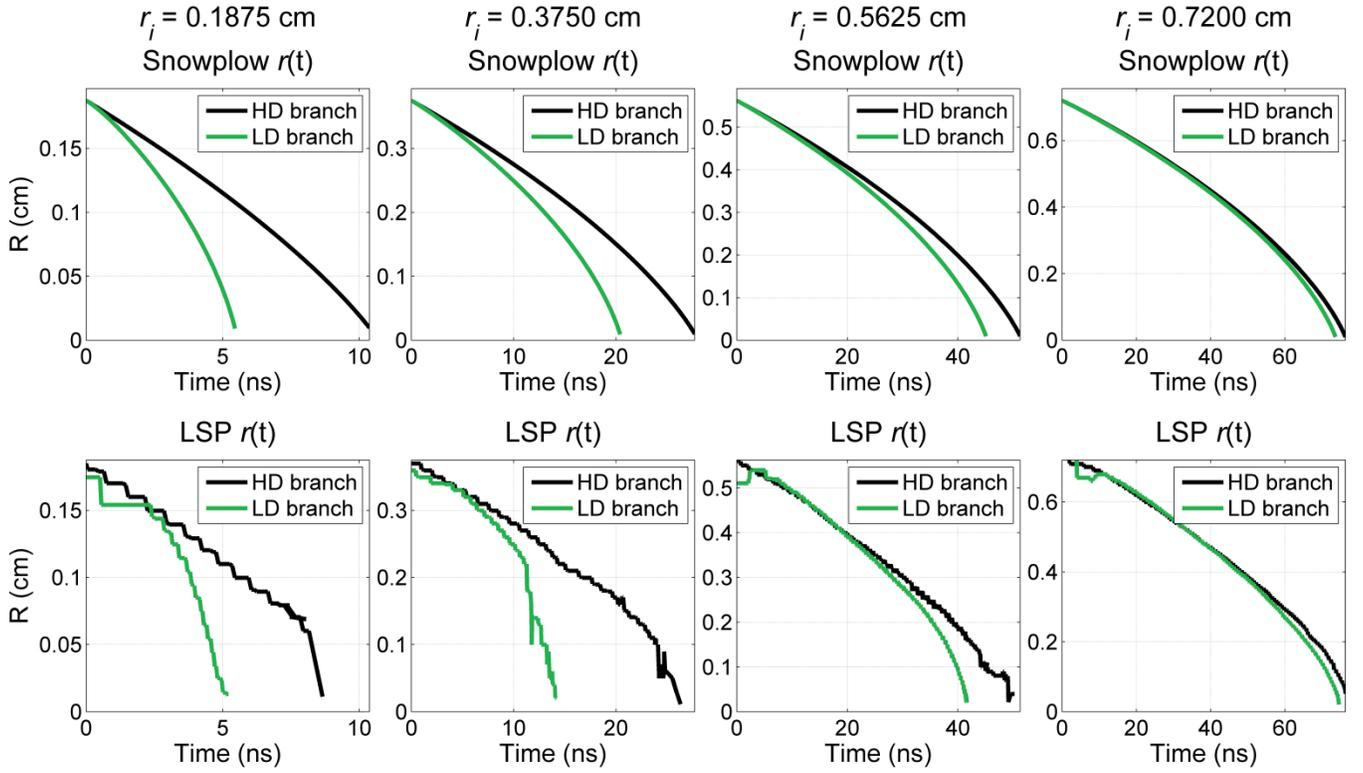

FIG. 10: The first row depicts the radial positions of the high density (HD) and low density (LD) slugs as calculated by the snowplow model for the different anode inner radii. The second row depicts the radial positions of the peak densities of the HD and LD regions of the sheath extracted from the LSP simulations. For a particular inner radius, the same horizontal and vertical scales are used for the LSP and the snowplow model results.

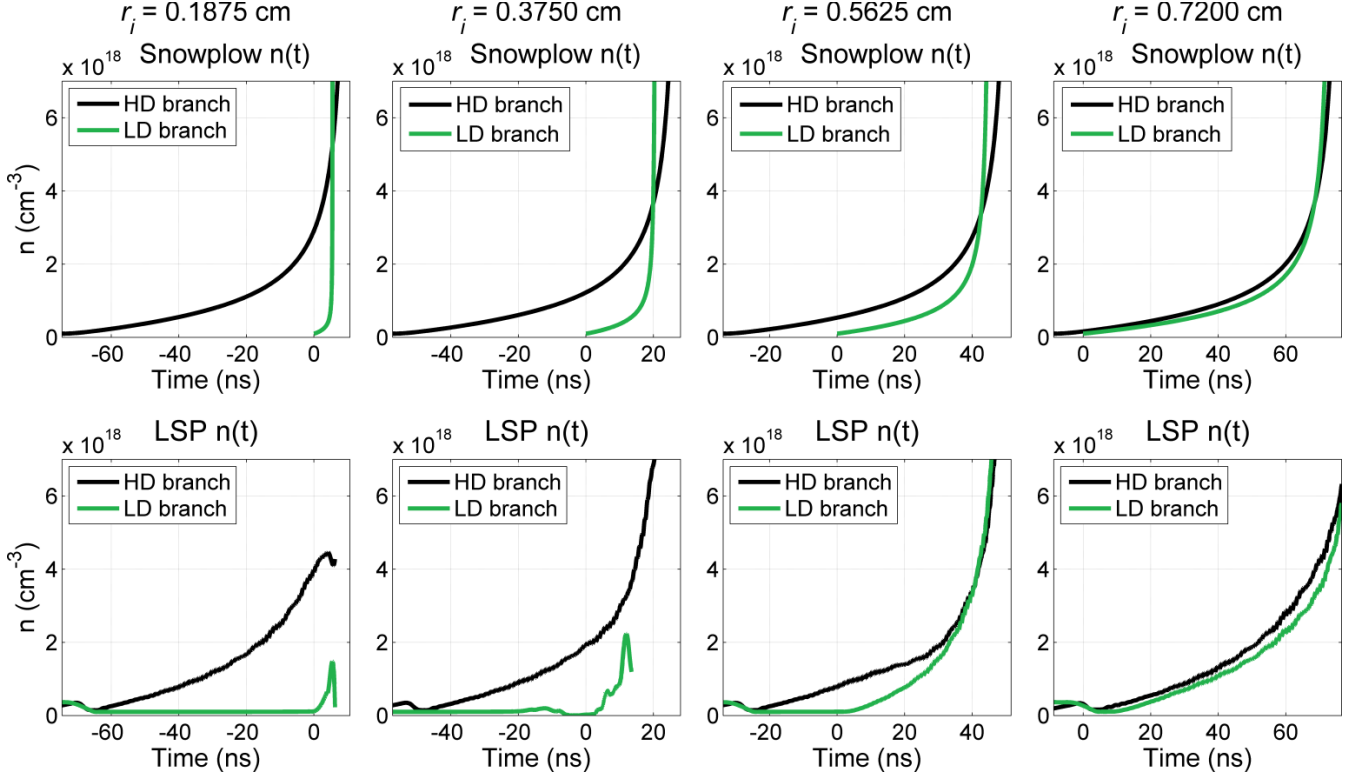

FIG. 11: The first row depicts the number density of the high density (HD) and low density (LD) slugs as calculated by the snowplow model for the different anode inner radii. The second row depicts the peak number density within the HD and LD regions of the sheath extracted from the LSP simulations. For a particular inner radius, the same horizontal and vertical scales are used for the LSP and the snowplow model results.

## 8. CONCLUSION

This paper introduces a way to significantly improve the neutron yield of the dense plasma focus. Fully kinetic simulations have revealed the short-lived existence of low density regions along the plasma sheath that form when the plasma passes a sharp corner of the anode. By introducing a hollow in the anode, such a low density region develops as the sheath passes the hollow radius during the radial implosion phase. As the sheath continues to implode, the plasma forms an axially non-uniform plasma annulus composed of a low density region alongside the high density bulk of the imploding plasma. This low density region only exists for a short duration, because the sheath rises in density by sweeping up background gas along its trajectory. Therefore, this effect is only significant for sufficiently small hollow inner radii: for too large of a radius, the low density region that forms has a significant amount of time to accumulate mass and build up to a density comparable to the remainder of the sheath. For anode hollows below a certain radius, the low density region that forms has no time to equalize in density with the rest of the sheath, and when it reaches the axis, it forms an axially non-uniform plasma column. Therefore, by selecting an appropriate inner radius, one can preferentially form a non-uniform plasma and thus seed the $m = 0$ instability to improve the ion beam generation and neutron yield. Thus far, kinetic simulations have demonstrated the enhanced $m = 0$ instability formation and improved neutron yield for sufficiently small anode hollow radii. The next step will be to experimentally verify such effects. Although trends in neutron yield will be straightforward to study, difficulty will lie in diagnosing the low density sheath formation, as the phenomenon is hidden within the anode itself.

A simple snowplow model has also been applied to calculate the density and radial trajectory of the plasma sheath during its implosion phase. This has yielded results that agree well with the kinetic simulations, and the next step will be to devise a method with which the snowplow model can determine the susceptibility of different anode geometries to seeding the $m = 0$ instability. However, some parameters used in the snowplow model, such as the snowplow thickness, are obtained from the kinetic simulations. Therefore, it is currently not possible to completely extricate the snowplow model from the kinetic simulations. Nonetheless, this is still a promising avenue towards a simple method of optimizing the anode inner radius for improved neutron yield without running time-consuming kinetic simulations: currently, a single kinetic simulation requires over 15k core-hours to complete whereas the snowplow model can be implemented on a traditional desktop computer and run in seconds.

**ACKNOWLEDGEMENTS**

This work was supported by the Department of Energy and Lawrence Livermore National Lab through Contract DE-AC52-07NA27344 the Laboratory Directed Research and Development Program (15-ERD-034).